\documentstyle[12pt]{article}

\begin{document}

\author{C. Bizdadea and S. O. Saliu\thanks{%
e-mail address: odile.saliu@comp-craiova.ro} \\
Department of Physics, University of Craiova\\
13 A. I. Cuza Str., Craiova R-1100, Romania}
\title{A Note on ``Irreducible'' $p$-Form Gauge Theories with Stueckelberg
Coupling }
\maketitle

\begin{abstract}
$p$-form gauge theories with Stueckelberg coupling are quantized in an
irreducible antifield-BRST way. As a consequence, neither the ghosts of
ghosts nor their antifields appear. Some irreducible gauge conditions are
inferred naturally within our formalism. In the end we briefly discuss the
interacting case.

\end{abstract}

The necessity of covariantly quantizing gauge theories with open algebras
has stimulated the development of the antifield-BRST formalism \cite{8}--%
\cite{12}, which represents the most powerful quantization method known so
far. The power of this treatment resides, among others, in its capacity of
being applied to reducible gauge theories, i.e., to those systems whose
gauge transformations are not independent. A typical class of redundant
models is expressed by gauge theories involving $p$-form gauge fields. Such
theories are important due to their link with string theory and supergravity
models \cite{1}--\cite{6}. Their BRST quantization \cite{17} has been
performed until now only in a reducible manner by introducing ghosts of
ghosts and a pyramid of non-minimal variables.

In this paper we prove that $p$-form gauge theories with Stueckelberg
coupling allow for an irreducible antifield-BRST quantization. As a
consequence of our irreducible treatment, the ghosts of ghosts and their
antifields do not appear. The main steps implied in our procedure consist
in: (i) the construction of an irreducible system associated with the
starting redundant model; (ii) the proof of the fact that we can
consistently replace the BRST quantization of the initial theory with the
quantization of the irreducible system; (iii) the antifield-BRST
quantization of the irreducible model. Finally, the analysis is extended to
interacting theories.

We start with the Lagrangian action \cite{18} 
\begin{eqnarray}\label{113}
& &S_0^L\left[ A_{\mu _1\ldots \mu _p},H_{\mu _1\ldots 
\mu _{p-1}}\right] =-\int d^Dx\left( \frac 1{2\cdot \left( p+1\right) !}
F_{\mu _1\ldots \mu _{p+1}}F^{\mu _1\ldots \mu _{p+1}}+
\right. \nonumber \\
& &\left. \frac 1{2\cdot p!}\left( MA_{\mu _1\ldots \mu _p}-
F_{\mu _1\ldots \mu _p}\right) \left( MA^{\mu _1\ldots \mu _p}-
F^{\mu _1\ldots \mu _p}\right) \right) , 
\end{eqnarray} where $F_{\mu _1\ldots \mu _{p+1}}$ and $F_{\mu _1\ldots \mu
_p}$ represent the field strengths of $A_{\mu _1\ldots \mu _p}$,
respectively, $H_{\mu _1\ldots \mu _{p-1}}$, and $M$ is a real constant. It
is understood that $D>p+1$. Action (\ref{113}) is invariant under the gauge
transformations $\delta _\epsilon \Phi ^{\alpha _0}=Z_{\;\;\alpha
_1}^{\alpha _0}\epsilon ^{\alpha _1}$, with $\Phi ^{\alpha _0}=\left( 
\begin{array}{c}
A^{\mu _1\ldots \mu _p} \\ 
H^{\mu _1\ldots \mu _{p-1}} 
\end{array}
\right) $, $\epsilon ^{\alpha _1}=\left( 
\begin{array}{c}
\epsilon ^{\nu _1\ldots \nu _{p-1}} \\ 
\bar \epsilon ^{\nu _1\ldots \nu _{p-2}} 
\end{array}
\right) $, and 
\begin{equation}
\label{116}Z_{\;\;\alpha _1}^{\alpha _0}=\left( 
\begin{array}{cc}
\frac 1{\left( p-1\right) !}\partial _{}^{\left[ \mu _1\right. }\delta
_{\;\;\nu _1}^{\mu _2}\ldots \delta _{\;\;\nu _{p-1}}^{\left. \mu _p\right]
} & {\bf 0} \\ \frac M{\left( p-1\right) !}\delta _{\;\;\nu _1}^{\left[ \mu
_1\right. }\ldots \delta _{\;\;\nu _{p-1}}^{\left. \mu _{p-1}\right] } & 
\frac 1{\left( p-2\right) !}\partial _{}^{\left[ \mu _1\right. }\delta
_{\;\;\nu _1}^{\mu _2}\ldots \delta _{\;\;\nu _{p-2}}^{\left. \mu
_{p-1}\right] } 
\end{array}
\right) , 
\end{equation}
The prior gauge transformations are $\left( p-1\right) $-stage reducible 
\begin{equation}
\label{118}Z_{\;\;\alpha _k}^{\alpha _{k-1}}Z_{\;\;\alpha _{k+1}}^{\alpha
_k}=0,\;k=1,\ldots p-1, 
\end{equation}
with the $k$th order reducibility functions of the form 
\begin{equation}
\label{119}Z_{\;\;\alpha _{k+1}}^{\alpha _k}=\left( 
\begin{array}{cc}
\frac 1{\left( p-k-1\right) !}\partial _{}^{\left[ \mu _1\right. }\delta
_{\;\;\nu _1}^{\mu _2}\ldots \delta _{\;\;\nu _{p-k-1}}^{\left. \mu
_{p-k}\right] } & {\bf 0} \\ \frac{\left( -\right) ^kM}{\left( p-k-1\right) !%
}\delta _{\;\;\nu _1}^{\left[ \mu _1\right. }\ldots \delta _{\;\;\nu
_{p-k-1}}^{\left. \mu _{p-k-1}\right] } & \frac 1{\left( p-k-2\right)
!}\partial _{}^{\left[ \mu _1\right. }\delta _{\;\;\nu _1}^{\mu _2}\ldots
\delta _{\;\;\nu _{p-k-2}}^{\left. \mu _{p-k-1}\right] } 
\end{array}
\right) , 
\end{equation}
where $\alpha _k=\left( \mu _1\ldots \mu _{p-k},\mu _1\ldots \mu
_{p-k-1}\right) $, $k=0,\ldots p-1$. Throughout this paper we work with the
conventions $f^{\mu _1\ldots \mu _m}=f$ if $m=0$, and $f^{\mu _1\ldots \mu
_m}=0$ if $m<0$.

Initially, we construct the irreducible theory associated with the starting
model. Corresponding to the relations (\ref{118}), we introduce the fields $%
\Phi ^{\alpha _{k+1}}\equiv \left( A^{\mu _1\ldots \mu _{p-k-1}},H^{\mu
_1\ldots \mu _{p-k-2}}\right) $ for every $k\geq 1$ odd, and the gauge
parameters $\epsilon ^{\alpha _{k+1}}\equiv \left( \epsilon ^{\mu _1\ldots
\mu _{p-k-1}},\bar \epsilon ^{\mu _1\ldots \mu _{p-k-2}}\right) $ for every $%
k\geq 2$ even. We take the gauge transformations of the new fields of the
form $\delta _\epsilon A^{\mu _1\ldots \mu _{p-2k-2}}=\partial ^{\left[ \mu
_1\right. }\epsilon ^{\left. \mu _2\ldots \mu _{p-2k-2}\right] }-M\bar
\epsilon ^{\mu _1\ldots \mu _{p-2k-2}}+\partial _\nu \epsilon ^{\nu \mu
_1\ldots \mu _{p-2k-2}}$ (with $k=\;0,\ldots ,a$), and $\delta _\epsilon
H^{\mu _1\ldots \mu _{p-2k-3}}=\;\partial ^{\left[ \mu _1\right. }\bar
\epsilon ^{\left. \mu _2\ldots \mu _{p-2k-3}\right] }+M\epsilon ^{\mu
_1\ldots \mu _{p-2k-3}}+\partial _\nu \bar \epsilon ^{\nu \mu _1\ldots \mu
_{p-2k-3}}$ (with $k=0,\ldots ,b$). We employed the notations $a=p/2-1$, $%
b=p/2-2$ for $p$ even, respectively, $a=\left( p-3\right) /2$ and $b=\left(
p-3\right) /2$ for $p$ odd. All the gauge transformations of the original
and newly introduced fields are irreducible. Indeed, taking $\epsilon ^{\mu
_1\ldots \mu _{p-2k-1}}=\partial ^{\left[ \mu _1\right. }\theta ^{\left. \mu
_2\ldots \mu _{p-2k-1}\right] }$ and $\bar \epsilon ^{\mu _1\ldots \mu
_{p-2k-2}}=\partial ^{\left[ \mu _1\right. }\bar \theta ^{\left. \mu
_2\ldots \mu _{p-2k-2}\right] }-M\theta ^{\mu _1\ldots \mu _{p-2k-2}}$, the
above gauge transformations vanish if and only if one has $\theta ^{\mu
_1\ldots \mu _{p-2k}}=\partial ^{\left[ \mu _1\right. }\xi ^{\left. \mu
_2\ldots \mu _{p-2k}\right] }$ and $\bar \theta ^{\mu _2\ldots \mu
_{p-2k-1}}=\partial ^{\left[ \mu _1\right. }\bar \xi ^{\left. \mu _2\ldots
\mu _{p-2k-1}\right] }-M\xi ^{\mu _1\ldots \mu _{p-2k-1}}$, hence $\epsilon
^{\mu _1\ldots \mu _{p-2k-1}}=0$ and $\bar \epsilon ^{\mu _1\ldots \mu
_{p-2k-2}}=0$, such that the irreducibility is manifest. Now, we consider
the theory described by the Lagrangian action $S_0^L[A^{\mu _1\ldots \mu
_{p-2k}},H^{\mu _1\ldots \mu _{p-2k-1}}]=S_0^L\left[ A^{\mu _1\ldots \mu
_p},H^{\mu _1\ldots \mu _{p-1}}\right] $, subject to the prior irreducible
gauge transformations. In this way, we associated an irreducible model with
the starting reducible theory. This irreducible system will be relevant by
virtue of the subsequent antifield-BRST analysis. It is well-known that the
BRST differential, $s$, splits into two differentials playing different
roles. The first differential, which is usually called the Koszul-Tate
operator, $\delta $, realizes an algebraic resolution of the smooth
functions defined on the stationary surface of field equations. The main
feature of this operator is its acyclicity, namely $H_l\left( \delta \right)
=0$ for every non-vanishing antighost number $l$, where $H_l\left( \delta
\right) $ denotes the $l$th order homological group of $\delta $. The second
differential is the (model of) longitudinal exterior derivative along the
gauge orbits, $D$, and takes into account the gauge invariances on the
stationary surface. In the sequel we analyze these two differentials.

First, we investigate the acyclicity of $\delta $ in connection with the
irreducible model built above. The minimal antifield spectrum includes the
variables $A_{\mu _1\ldots \mu _{p-2k}}^{*}$, $k=0,\ldots ,c$, $H_{\mu
_1\ldots \mu _{p-2k-1}}^{*}$, $k=0,\ldots ,d$, $\eta _{\mu _1\ldots \mu
_{p-2k-1}}^{*}$, $k=0,\ldots ,d$, and $C_{\mu _1\ldots \mu _{p-2k-2}}^{*}$, $%
k=0,\ldots ,a$, where $c=p/2$, $d=p/2-1$ for $p$ even, respectively, $%
c=d=\left( p-1\right) /2$ for $p$ odd. The antifields $A^{*}$ and $H^{*}$
are fermionic and of antighost number one, while the antifields $\eta ^{*}$
and $C^{*}$ are all bosonic and possess antighost number two. The standard
BRST definitions of $\delta $ acting on the fields and antighost number one
antifields yield $\delta A^{\mu _1\ldots \mu _{p-2k}}=0$, $k=0,\ldots ,c$, $%
\delta H^{\mu _1\ldots \mu _{p-2k-1}}=0$, $k=0,\ldots ,d$, $\delta A_{\mu
_1\ldots \mu _p}^{*}=-\left( 1/p!\right) \partial ^\nu F_{\nu \mu _1\ldots
\mu _p}+\left( M/p!\right) \left( MA_{\mu _1\ldots \mu _p}-F_{\mu _1\ldots
\mu _p}\right) $, $\delta H_{\mu _1\ldots \mu _{p-1}}^{*}=-\,\left( 1/\left(
p-1\right) !\right) \,\partial ^\nu F_{\nu \mu _1\ldots \mu _{p-1}}\;+\left(
M/\left( p-1\right) !\right) \partial ^\nu A_{\nu \mu _1\ldots \mu _{p-1}}$, 
$\delta A_{\mu _1\ldots \mu _{p-2k}}^{*}=0$, $k=1,\ldots ,c$, $\delta H_{\mu
_1\ldots \mu _{p-2k-1}}^{*}=0$, $k=1,\ldots ,d$, while $\delta $ acts on the
antighost number two antifields through 
\begin{eqnarray}\label{5}
& &\delta \eta _{\mu _1\ldots \mu _{p-2k-1}}^{*}=
-\left( p-2k\right) \partial ^\mu A_{\mu \mu _1\ldots 
\mu _{p-2k-1}}^{*}+MH_{\mu _1\ldots 
\mu _{p-2k-1}}^{*}- \nonumber \\
& &\frac 1{p-2k-1}\partial _{\left[ \mu _1\right. }
A_{\left. \mu _2\ldots \mu _{p-2k-1}\right] }^{*},\;k=0,\ldots ,d, 
\end{eqnarray} 
\begin{eqnarray}\label{6}
& &\delta C_{\mu _1\ldots \mu _{p-2k-2}}^{*}=
-\left( p-2k-1\right) \partial ^\mu H_{\mu \mu _1\ldots 
\mu _{p-2k-2}}^{*}-MA_{\mu _1\ldots 
\mu _{p-2k-2}}^{*}- \nonumber \\
& &\frac 1{p-2k-2}\partial _{\left[ \mu _1\right. }
H_{\left. \mu _2\ldots \mu _{p-2k-2}\right] }^{*},\;k=0,\ldots ,a. 
\end{eqnarray} We prove that with the present antifield spectrum at hand,
the acyclicity of $\delta $ is ensured on the basis of the prior
definitions. From the above definitions, we find at resolution degree one
the non-trivial co-cycles $\partial ^\mu H_{\mu \mu _1\ldots \mu _{p-2}}^{*}$%
, $u_{\mu _1\ldots \mu _{p-1}}\equiv \partial ^\mu A_{\mu \mu _1\ldots \mu
_{p-1}}^{*}-\frac MpH_{\mu _1\ldots \mu _{p-1}}^{*}$, $A_{\mu _1\ldots \mu
_{p-2k}}^{*}$, $k=1,\ldots ,c$, and $H_{\mu _1\ldots \mu _{p-2k-1}}^{*}$, $%
k=1,\ldots ,d$. We show that all these co-cycles are also $\delta $-exact.
The proof is given for definiteness in the case $p$ even, the other
situation being similar. The starting point is represented by the last
equations from (\ref{5}) and (\ref{6}), namely $\delta \eta _\mu
^{*}=-2\partial ^\nu A_{\nu \mu }^{*}+MH_\mu ^{*}-\partial _\mu A^{*}$,
respectively, $\delta C^{*}=-\partial ^\mu H_\mu ^{*}-MA^{*}$. Applying $%
\partial ^\mu $ on the first from the above relations, and using the other
one, we find $A^{*}=\delta \left( -\frac 1{\Box +M^2}\left( \partial ^\mu
\eta _\mu ^{*}+MC^{*}\right) \right) $. Taking into account the next
relation from (\ref{6}), we derive in the same manner that $H_\mu
^{*}=\delta \left( -\frac 1{\Box +M^2}\left( 2\partial ^\nu C_{\nu \mu
}^{*}-M\eta _\mu ^{*}+\partial _\mu C^{*}\right) \right) $. Step by step, we
get 
\begin{eqnarray}\label{7}
& &A_{\mu _1\ldots \mu _{p-2k-2}}^{*}=\delta \left( -
\frac 1{\Box +M^2}\left( \left( p-2k-1\right) \partial ^\mu 
\eta _{\mu \mu _1\ldots \mu _{p-2k-2}}^{*}+MC_{\mu _1\ldots 
\mu _{p-2k-2}}^{*}\right. \right. \nonumber \\
& &\left. \left. +\frac 1{p-2k-2}\partial _{\left[ \mu _1\right. }
\eta _{\left. \mu _2\ldots \mu _{p-2k-2}
\right] }^{*}\right) \right) ,\;k=0,\ldots ,a, 
\end{eqnarray} 
\begin{eqnarray}\label{8}
& &H_{\mu _1\ldots \mu _{p-2k-3}}^{*}=\delta \left( -
\frac 1{\Box +M^2}\left( \left( p-2k-2\right) \partial ^\mu 
C_{\mu \mu _1\ldots \mu _{p-2k-3}}^{*}-M\eta _{\mu _1
\ldots \mu _{p-2k-3}}^{*}\right. \right. \nonumber \\
& &\left. \left. +\frac 1{p-2k-3}\partial _{\left[ \mu _1\right. }
C_{\left. \mu _2\ldots \mu _{p-2k-3}\right] }^{*}\right) 
\right) ,\;k=0,\ldots ,b, 
\end{eqnarray} 
\begin{equation}
\label{9}\partial ^\mu H_{\mu \mu _1\ldots \mu _{p-2}}^{*}=\delta \left( 
\frac{\partial ^\mu }{\Box +M^2}\left( M\eta _{\mu \mu _1\ldots \mu
_{p-2}}^{*}-\frac 1{p-1}\partial _{\left[ \mu \right. }C_{\left. \mu
_1\ldots \mu _{p-2}\right] }^{*}\right) \right) , 
\end{equation}
\begin{eqnarray}\label{10}
& &u_{\mu _1\ldots \mu _{p-1}}=\delta \left( -
\frac 1{p\left( \Box +M^2\right) }\left( M^2\eta _{\mu _1
\ldots \mu _{p-1}}^{*}+\partial ^\mu 
\partial _{\left[ \mu \right. }\eta _{\left. \mu _1\ldots 
\mu _{p-1}\right] }^{*}\right. \right. \nonumber \\
& &\left. \left. -\frac M{p-1}\partial _{\left[ \mu _1\right. }
C_{\left. \mu _2\ldots \mu _{p-1}\right] }^{*}\right) \right) , 
\end{eqnarray} indicating that all the above mentioned co-cycles are trivial
in the homology of $\delta $. With these results at hand, and, at the same
time, invoking the irreducibility of the gauge transformations, it follows
that there are no other non-trivial co-cycles, so $\delta $ is acyclic. We
underline that the acyclicity was established employing just the previous
antifield spectrum, that contains only generators of antighost number one or
two.

Next, we investigate the longitudinal exterior derivative along the gauge
orbits associated with the irreducible model. The minimal ghost spectrum
contains the fermionic pure ghost number one ghost fields $\eta ^{\mu
_1\ldots \mu _{p-2k-1}}$, $k=0,\ldots ,d$ and $C^{\mu _1\ldots \mu
_{p-2k-2}} $, $k=0,\ldots ,a$. The usual definition of $D$ acting on the
fields leads to $DA^{\mu _1\ldots \mu _p}=\partial ^{\left[ \mu _1\right.
}\eta ^{\left. \mu _2\ldots \mu _p\right] }$, $DH^{\mu _1\ldots \mu
_{p-1}}=\partial ^{\left[ \mu _1\right. }C^{\left. \mu _2\ldots \mu
_{p-1}\right] }+M\eta ^{\mu _1\ldots \mu _{p-1}}$, $DA^{\mu _1\ldots \mu
_{p-2k-2}}=\partial ^{\left[ \mu _1\right. }\eta ^{\left. \mu _2\ldots \mu
_{p-2k-2}\right] }-MC^{\mu _1\ldots \mu _{p-2k-2}}+\partial _\nu \eta ^{\nu
\mu _1\ldots \mu _{p-2k-2}}$ (with $k=0,\ldots ,a$), $DH^{\mu _1\ldots \mu
_{p-2k-3}}=\partial ^{\left[ \mu _1\right. }C^{\left. \mu _2\ldots \mu
_{p-2k-3}\right] }+M\eta ^{\mu _1\ldots \mu _{p-2k-3}}+\partial _\nu C^{\nu
\mu _1\ldots \mu _{p-2k-3}}$ (with $k=0,\ldots ,b$), while for the ghosts we
standardly have $D\eta ^{\mu _1\ldots \mu _{p-2k-1}}=0$ (for $k=0,\ldots ,d$%
), $DC^{\mu _1\ldots \mu _{p-2k-2}}=0$ (for $k=0,\ldots ,a$). First, we show
that the irreducible and original models possess the same observables. The
equations fulfilled by an observable, $F$, of the irreducible model read as 
\begin{equation}
\label{160a}m^{(k)}\partial _{\left[ \mu _1\right. }\frac{\delta F}{\delta
H^{\left. \mu _2\ldots \mu _{p-2k-2}\right] }}+M\frac{\delta F}{\delta
A^{\mu _1\ldots \mu _{p-2k-2}}}+n^{(k)}\partial ^\mu \frac{\delta F}{\delta
H^{\mu \mu _1\ldots \mu _{p-2k-2}}}=0, 
\end{equation}
\begin{equation}
\label{160b}p^{(k)}\partial _{\left[ \mu _1\right. }\frac{\delta F}{\delta
A^{\left. \mu _2\ldots \mu _{p-2k-1}\right] }}-M\frac{\delta F}{\delta
H^{\mu _1\ldots \mu _{p-2k-1}}}+q^{(k)}\partial ^\mu \frac{\delta F}{\delta
A^{\mu \mu _1\ldots \mu _{p-2k-1}}}=0, 
\end{equation}
where $m^{(k)}=1/\left( p-2k-2\right) $, $n^{(k)}=p-2k-1$ for $k=0,\ldots a$%
, and $p^{(k)}=1/n^{(k)}$, $q^{(k)}=p-2k$ for $k=0,\ldots d$. We exploit
equations (\ref{160a}--\ref{160b}) level by level, beginning with the last
relations. For definiteness, we illustrate the procedure for $p$ even, the
opposite situation being treated along the same line. In this case, the last
equations are expressed by $M\frac{\delta F}{\delta A}+\partial ^\mu \frac{%
\delta F}{\delta H^\mu }=0$, and $\partial _\mu \frac{\delta F}{\delta A}-M%
\frac{\delta F}{\delta H^\mu }+2\partial ^\nu \frac{\delta F}{\delta A^{\nu
\mu }}=0$. Applying $\partial ^\mu $ on the second relation and using the
first equation, we find $\left( \partial ^\mu \partial _\mu +M^2\right) 
\frac{\delta F}{\delta A}=0$, which further implies $\frac{\delta F}{\delta A%
}=0$ due to the invertibility of the Klein-Gordon operator. Introducing $%
\frac{\delta F}{\delta A}=0$ in the starting relations, we obtain in
addition $\partial ^\mu \frac{\delta F}{\delta H^\mu }=0$ and $\partial ^\nu 
\frac{\delta F}{\delta A^{\nu \mu }}=\frac 12M\frac{\delta F}{\delta H^\mu }$%
. Taking into account the next equation from (\ref{160a}), namely $\frac
12\partial _{\left[ \mu \right. }\frac{\delta F}{\delta H^{\left. \nu
\right] }}+M\frac{\delta F}{\delta A^{\mu \nu }}+3\partial ^\rho \frac{%
\delta F}{\delta H^{\rho \mu \nu }}=0$, multiplied by $\partial ^\mu $ and
on behalf of the prior additional relations, it follows $\frac{\delta F}{%
\delta H^\mu }=0$. Similarly, we derive that $F$ does not depend on the new
fields, and, moreover, satisfies the equations $\partial ^\mu \frac{\delta F%
}{\delta H^{\mu \mu _1\ldots \mu _{p-2}}}=0$ and $M\frac{\delta F}{\delta
H^{\mu _1\ldots \mu _{p-1}}}-p\partial ^\mu \frac{\delta F}{\delta A^{\mu
\mu _1\ldots \mu _{p-1}}}=0$, which are nothing but the equations checked by
any observable of the redundant theory. Thus, any observable of the
irreducible system is an observable of the reducible one. The converse also
holds, i.e., any observable, $\bar F$, of the reducible model is an
observable of the irreducible one. This is as $\bar F$ clearly verifies (\ref
{160a}--\ref{160b}). Accordingly the ingredients of the antifield-BRST
formalism, the zeroth order cohomological groups corresponding to the
reducible, respectively, irreducible longitudinal exterior derivatives along
the gauge orbits coincide. In agreement with the homological perturbation
theory \cite{13}--\cite{16}, the zeroth order cohomological groups
associated with the reducible and irreducible BRST symmetries are equal. At
the same time, the acyclicity of the irreducible Koszul-Tate operator
ensures the nilpotency of the corresponding BRST symmetry. These results
make permissible from the physical point of view, i.e., from the point of
view of the requirements $s^2=0$ and $H^0(s)=\left\{ {\rm observables}%
\right\} $, the replacement of the BRST quantization of the reducible model
with the one of the irreducible system constructed above. $H^0(s)$ denotes
the zeroth order cohomological group of $s$.

Although the cohomological groups of the longitudinal exterior derivatives
at pure ghost number zero coincide, this may not hold at superior pure
ghost numbers. Indeed, in the irreducible case the higher order
cohomological groups of $D$ vanish as the ghosts are all $D$-exact. This can
be shown as follows for $p$ even (the opposite situation is similar). We
begin with the last equation from the expressions of $D$ acting on the
fields, namely, $DA=-MC+\partial _\mu \eta ^\mu $, $DH^\mu =\partial ^\mu
C+M\eta ^\mu +\partial _\nu C^{\nu \mu }$. Applying $\partial ^\mu $ on the
latter and using the former, we find $C=D\left( \frac 1{\Box +M^2}\left(
\partial _\mu H^\mu -MA\right) \right) $. Inserting the prior expression of $%
C$ in the starting equations and applying $\partial ^\mu $ on the relation
expressing $DA^{\mu \nu }$, we derive $\eta ^\mu =D\left( \frac 1{\Box
+M^2}\left( \partial ^\mu A+MH^\mu +\partial _\nu A^{\nu \mu }\right)
\right) $. Acting along the same line, we output $\eta ^{\mu _1\ldots \mu
_{p-2k-1}}=D\left( \frac 1{\Box +M^2}\chi ^{\mu _1\ldots \mu
_{p-2k-1}}\right) $ (with $k=0,\ldots ,d$), and $C^{\mu _1\ldots \mu
_{p-2k-2}}=D\left( \frac 1{\Box +M^2}\varphi ^{\mu _1\ldots \mu
_{p-2k-2}}\right) $ (with $k=0,\ldots ,a$), where $\chi ^{\mu _1\ldots \mu
_{p-2k-1}}=\partial ^{\left[ \mu _1\right. }A^{\left. \mu _2\ldots \mu
_{p-2k-1}\right] }+MH^{\mu _1\ldots \mu _{p-2k-1}}+\partial _\nu A^{\nu \mu
_1\ldots \mu _{p-2k-1}}$, and $\varphi ^{\mu _1\ldots \mu
_{p-2k-2}}=\partial ^{\left[ \mu _1\right. }H^{\left. \mu _2\ldots \mu
_{p-2k-2}\right] }-MA^{\mu _1\ldots \mu _{p-2k-2}}+\partial _\nu H^{\nu \mu
_1\ldots \mu _{p-2k-2}}$. Because the ghosts are all $D$-closed, it results
that any $D$-closed quantity with pure ghost number greater than zero is a
polynomial in the ghosts with coefficients which are gauge invariant
functions. On the other hand, as all the ghosts are $D$-exact, we have that
any such $D$-closed quantity is also $D$-exact, so all higher order
cohomological groups of $D$ vanish in the irreducible case. In the reducible
situation, however, not necessarily all the higher order cohomological groups 
are trivial. The analysis of the cohomology of the irreducible $D$ clearly
outlines that the functions $\chi ^{\mu _1\ldots \mu _{p-2k-1}}$ and $%
\varphi ^{\mu _1\ldots \mu _{p-2k-2}}$ may be regarded as some purely gauge
fields with the gauge transformations $\delta _\omega \chi ^{\mu _1\ldots
\mu _{p-2k-1}}=\left( \Box +M^2\right) \omega ^{\mu _1\ldots \mu _{p-2k-1}}$%
, $\delta _{\bar \omega }\varphi ^{\mu _1\ldots \mu _{p-2k-2}}=\bar \omega
^{\mu _1\ldots \mu _{p-2k-2}}$, so it is natural to choose $\chi ^{\mu
_1\ldots \mu _{p-2k-1}}=0$ and $\varphi ^{\mu _1\ldots \mu _{p-2k-2}}=0$
like gauge conditions (because they are in addition irreducible). Hence, our
formalism displays a class of possible gauge conditions that can be used at
the BRST quantization during the gauge-fixing process.

We discussed before that it is legitimate to substitute the BRST
quantization of the reducible theory with the quantization of the
irreducible model. In the sequel we pass to the antifield-BRST quantization
of the irreducible system. With the ghost and antifield spectra introduced
previously, we take the non-minimal sector $\left( \bar \eta ^{\mu _1\ldots
\mu _{p-2k-1}},\bar \eta _{\mu _1\ldots \mu _{p-2k-1}}^{*}\right) $, $\left(
B^{\mu _1\ldots \mu _{p-2k-1}},B_{\mu _1\ldots \mu _{p-2k-1}}^{*}\right) $,
for $k=0,\ldots ,d$ and $\left( \bar C^{\mu _1\ldots \mu _{p-2k-2}},\bar
C_{\mu _1\ldots \mu _{p-2k-2}}^{*}\right) $, $\left( {\cal B}^{\mu _1\ldots
\mu _{p-2k-2}},{\cal B}_{\mu _1\ldots \mu _{p-2k-2}}^{*}\right) $, for $%
k=0,\ldots ,a$, with the $B$'s, ${\cal B}$'s, $\bar \eta ^{*}$'s and $\bar
C^{*}$'s bosonic, with ghost number zero, and the remaining fields
fermionic, with ghost number minus one (the ghost number is difference
between the pure ghost and antighost numbers). Consequently, the non-minimal
solution of the master equation is given by 
\begin{eqnarray*}
& &S=S_0^L+\int d^Dx\left( A_{\mu _1\ldots \mu _p}^
{*}\partial ^{\left[ \mu _1\right. }\eta ^{\left. \mu _2\ldots 
\mu _p\right] }+\sum_{k=0}^d\bar \eta _{\mu _1\ldots 
\mu _{p-2k-1}}^{*}B^{\mu _1\ldots \mu _{p-2k-1}}+\right. \\
& &\sum_{k=0}^aA_{\mu _1\ldots \mu _{p-2k-2}}^{*}\left( \partial ^
{\left[ \mu _1\right. }\eta ^{\left. \mu _2\ldots \mu _{p-2k-2}\right] }-
MC^{\mu _1\ldots \mu _{p-2k-2}}+\partial _\mu \eta ^
{\mu \mu _1\ldots \mu _{p-2k-2}}\right) + \\
& &H_{\mu _1\ldots \mu _{p-1}}^{*}\left( M\eta ^{\mu _1\ldots 
\mu _{p-1}}+\partial ^{\left[ \mu _1\right. }C^{\left. \mu _2\ldots 
\mu _{p-1}\right] }\right) +\sum_{k=0}^a\bar C_{\mu _1\ldots 
\mu _{p-2k-2}}^{*}B^{\mu _1\ldots \mu _{p-2k-2}}+ \\
& &\left. \sum_{k=0}^bH_{\mu _1\ldots \mu _{p-2k-3}}^{*}\left( 
\partial ^{\left[ \mu _1\right. }C^{\left. \mu _2\ldots \mu _
{p-2k-3}\right] }+M\eta ^{\mu _1\ldots \mu _{p-2k-3}}+
\partial _\mu C^{\mu \mu _1\ldots \mu _{p-2k-3}}\right) \right) . 
\end{eqnarray*} In agreement with the above discussion, we choose the
gauge-fixing fermion $\psi =\int d^Dx\left( \sum_{k=0}^d\bar \eta _{\mu
_1\ldots \mu _{p-2k-1}}\chi ^{\mu _1\ldots \mu _{p-2k-1}}+\sum_{k=0}^a\bar
C_{\mu _1\ldots \mu _{p-2k-2}}\varphi ^{\mu _1\ldots \mu _{p-2k-2}}\right) $%
. The gauge-fixed action will be expressed by 
\begin{eqnarray}\label{167}
& &S_{\psi }=S_0^L+\int d^Dx\left( \sum_{k=0}^d\bar 
\eta _{\mu _1\ldots \mu _{p-2k-1}}\left( \Box +M^2\right) 
\eta ^{\mu _1\ldots \mu _{p-2k-1}}+\right. \nonumber \\
& &\sum_{k=0}^dB_{\mu _1\ldots \mu _{p-2k-1}} 
\chi ^{\mu _1\ldots \mu _{p-2k-1}}+\sum_{k=0}^a
{\cal{B}}_{\mu _1\ldots \mu _{p-2k-2}} 
\varphi ^{\mu _1\ldots \mu _{p-2k-2}}+ \nonumber \\
& &\left. +\sum_{k=0}^a\bar C_{\mu _1\ldots 
\mu _{p-2k-2}}\left( \Box +M^2\right) C^{\mu _1\ldots 
\mu _{p-2k-2}}\right) . 
\end{eqnarray}  It is clear that action (\ref{167}) possesses no residual
gauge invariances. Thus, we succeeded in quantizing the model with abelian $%
p $- and $\left( p-1\right) $-form gauge fields coupled through a
Stueckelberg-like term by employing an irreducible treatment, which needs no
ghosts of ghosts.

The results exposed here can be also applied to interacting theories. It is
known \cite{19} that the consistent interactions of a given gauge theory can
be such that: (a) they do not change the original gauge transformations; (b)
they change the gauge transformations, but do not afflict their gauge
algebra, or (c) they simultaneously change the gauge transformations, as
well as the gauge algebra. The case (a) can be directly approached within
the context of the procedure exposed in this paper. Indeed, if we add to (%
\ref{113}) an interaction term $S_I\left[ A_{\mu _1\ldots \mu _p},H_{\mu
_1\ldots \mu _{p-1}}\right] $ invariant under the original gauge
transformations, the previous analysis remains valid. The only modification
appears in relation with the definitions of $\delta $ acting on the
antifields associated with the original fields because the interaction term
may add new pieces to the field equations. This fact, however, does not
change in any way the prior treatment as the Noether identities of the
interacting theory are same with those of the model described by action (\ref
{113}). Then, the resulting gauge-fixed action of the interacting system
will be expressed by the right-hand side of (\ref{167}) to which one must
add $S_I\left[ A_{\mu _1\ldots \mu _p},H_{\mu _1\ldots \mu _{p-1}}\right] $.
The more realistic cases (b) and (c) (including the interactions that
satisfy the Durand-Weinberg-Witten theorem) can be also approached in an
irreducible manner, but in this situation it is necessary to apply the
technique of the consistent deformation of the master equation \cite{20}.
This represents a separate matter, which will be reported elsewhere. This
completes our approach. We remark that in the limit $M=0$ our procedure
leads to the irreducible quantization of a system with free abelian $p$- and 
$\left( p-1\right) $-form gauge fields.

To conclude with, in this paper we proved that $p$-form gauge theories with
Stueckelberg coupling can be consistently quantized within the irreducible
antifield-BRST formalism. The key point of our treatment consists in
constructing an irreducible theory in a way that makes permissible the
replacement of the redundant BRST quantization with the irreducible one.
Both the irreducible Koszul-Tate operator and the longitudinal exterior
derivative have been explicitly built. The analysis of the longitudinal
exterior derivative emphasizes some irreducible gauge conditions that can be
used at the gauge-fixing procedure. Finally, we mention that our results
open a new perspective for a cohomological approach to $p$-form gauge
theories.

\end{document}